\documentstyle[12pt]{article}

\def\lsim{\raise0.3ex\hbox{$\;<$\kern-0.75em\raise-1.1ex\hbox{$\sim\;$}}}
\def\gsim{\raise0.3ex\hbox{$\;>$\kern-0.75em\raise-1.1ex\hbox{$\sim\;$}}}

\def\beq{\begin{equation}}
\def\eeq{\end{equation}}
\def\barr{\begin{eqnarray}}
\def\earr{\end{eqnarray}}

\begin{document}

\rightline{hep-ph/9903329}
\bigskip

\begin{center}

{\Large {\bf 
Coherence and the Day - Night Asymmetry \\
in the Solar Neutrino Flux}}

\bigskip

{\large {\it Amol S. Dighe$^1$, Q. Y. Liu$^1$, 
        Alexei Yu. Smirnov$^{1,2}$}}

\medskip

{$^1$ {\it The Abdus Salam International Center for
Theoretical Physics, \\ Strada Costiera 11,  34100, Trieste,
Italy.}}

\smallskip

{$^2${\it Institute for Nuclear Research, RAN,
Moscow, Russia.}}

\end{center}

\bigskip

\begin{abstract}

We consider 
the day-night asymmetries predicted by  MSW solutions
of the solar neutrino problem. The integration 
over the neutrino energy, as well as over the production region
or over the time intervals of more than a day leads
to the averaging of oscillations on the way to the earth.
This is equivalent to considering 
the neutrino state arriving
at the surface of the earth as an {\it incoherent} mixture of the
neutrino {\it mass} eigenstates (even if there is no divergence
of wavepackets).
As a consequence, the $\nu_e$-regeneration effect
inside the earth is incoherent, in contrast with the result in 
hep-ph/9902435.

\end{abstract}

\medskip

{\bf 1.}
The day-night effect -- difference in the $\nu_e$ fluxes
during
the day and the night due to the earth matter -- is one of
the key signatures of the MSW solutions of the solar
neutrino problem \cite{ms86}. (The effect is negligible for the vacuum
oscillation solution.) Moreover, the zenith angle distribution
of the events during the night is different for the large 
mixing angle (LMA) and the small mixing angle (SMA) solutions.
(For recent discussions, see \cite{daynight}.)
Therefore, the day-night asymmetry and
the zenith angle dependence of the flux, if established,
will not only 
prove the MSW solution, but also distinguish among the different
MSW solutions.

Existing Super-Kamiokande (SK) results on the day-night
asymmetry and the zenith angle dependence \cite{sk1}
already exclude a significant range of the parameters
$\Delta m^2$ and $\sin^2 2\theta$ of the LMA solution and some part
of the SMA solution \cite{bks, sk1}. 
SK results after 708 days show a 6\%
excess of the 
number of events during the night  \cite{sk2}. 
This indication of the day-night effect is 
not yet statistically significant ($1.3 \sigma$).

The earth matter effect has been calculated, taking into 
account the divergence of the neutrino wave packet \cite{ms87} and the 
averaging effects \cite{daynight,ms87,baltz}. 
The neutrino state arriving
at the earth was considered to be an incoherent mixture
of the mass eigenstates $\nu_1$ and $\nu_2$.
The calculations give specific 
zenith angle ($\Theta_z$) 
distributions of the number of events during
the night, depending on the solution. In the case of the 
SMA solution, the excess during the night
is expected to be in the 
largest zenith angle 
bin $\cos \Theta_z = -1.0 \div -0.8$,
when the neutrinos cross both the mantle 
and the core of the earth \cite{daynight,rosen}. 
In the case of the LMA solution,
the excess of events will be distributed nearly uniformly
in all the night bins. Existing SK data \cite{sk1}
shows that the
excess is not concentrated in the largest $\Theta_z$ bin,
but is distributed in all the bins, thus indicating
towards the LMA solution.

In a recent paper \cite{sehgal}, 
the beam of neutrinos arriving at the earth 
was taken to be in a coherent state
so that the phase between the flavor eigenstates
is independent of the neutrino energy, the
point of production, or the time of the year.
This leads to a significant (a factor of 6)
enhancement of the day-night 
asymmetry and to the modification of
the zenith (nadir) angle distribution of the events. 
(For the SMA solution, the distribution is predicted
to be more uniform, with a wide maximum in the
bins $\cos \Theta_z = -0.8 \div -0.4$.)
Clearly, this would significantly change the implications
of the SK data.

In this paper, we show that due to averaging, the coherence
is effectively lost and the results of \cite{sehgal} are
not valid.

The need to take the averaging effects into account
has already been pointed out in \cite{ms87,baltz,lisi}. Here we shall
explicitly show where and how 
the averaging occurs and discuss
the conditions that need to be satisfied in order to observe
the coherence effects, at least in principle.

\bigskip

{\bf 2.}
For simplicity, let us consider the 2-neutrino mixing case:
\beq
\nu_e = \cos \theta ~\nu_1 + \sin \theta ~\nu_2 \; , \;
\nu_\mu = \sin \theta ~\nu_1 - \cos \theta ~\nu_2 ~~,
\label{eigendef}
\eeq 
where $\theta$ is the mixing angle in vacuum. 
(We will also use the notation 
$c \equiv \cos \theta$ and
$s \equiv \sin \theta$.)
In general, the neutrino state at the surface of the sun can
be written in terms of neutrino mass eigenstates as
\beq
\nu_\odot = a_1 ~\nu_1 + a_2 ~\nu_2 ~e^{i\phi_S}~~,
\label{nus}
\eeq
where $a_1$ and $a_2$ are the absolute values of the
amplitudes of $\nu_e \to \nu_1$ and $\nu_e \to \nu_2$
transitions inside the sun respectively.
They can be expressed in terms of $P_\odot \equiv
\overline{P(\nu_e \to \nu_e)}$, the averaged survival probability of
$\nu_e$ inside the sun. Indeed, according to
(\ref{eigendef}) and (\ref{nus}),
the amplitude of probability
to find $\nu_e$ at the surface of the sun equals
\beq
\langle \nu_e | \nu_\odot \rangle =
a_1 \cos \theta + a_2 \sin \theta e^{i\phi_S} ~~.
\eeq
Then
\beq
P_\odot \equiv \overline{|\langle \nu_e | \nu_\odot \rangle |^2}
= a_1^2 \cos^2 \theta + a_2^2 \sin^2 \theta ~~,
\label{psun}
\eeq
where averaging over the phase $\phi_S$ has been done. 
>From (\ref{psun}), we find 
\beq
a_2^2 \equiv 1 - a_1^2 = \frac{\cos^2 \theta - P_\odot}{\cos 2\theta}~~.
\label{a2def}
\eeq

Let us assume that there is no loss of coherence 
due to the spread of the wave packet on the way to the
earth. 
(This loss of coherence would further dilute
the interference effects
\cite{ms87, sehgal}, but we shall not consider it here.)
Then, on the way from the sun to the earth,
the state (\ref{nus})  evolves to
\beq
\nu_E = a_1 ~\nu_1 + a_2 ~\nu_2  ~e^{i(\phi_S + \varphi)}~~,
\eeq
where 
\beq
\varphi = 2 \pi \frac{L}{\ell_\nu} = \frac{L ~ \Delta m^2}{2 E}
\label{phase}
\eeq
is the (relative) phase acquired by the mass eigenstates on the 
way from the sun to the earth. Here
$L \approx 1.5 \times 10^{13}$ cm
is the distance between 
the sun and the earth.

The key point is that, for the MSW solutions of the
solar neutrino problem, the distance $L$ is much greater
than the oscillation length $\ell_\nu$. Indeed, for
the SMA solution characterized by
$\Delta m^2 \approx 5 \times 10^{-6}$ eV$^2$, and for
the typical energy of Boron neutrinos ($E \sim 10$ MeV), 
we get $\ell_\nu \approx 5 \times 10^8$ cm, so that 
\beq
\frac{\ell_\nu}{L} 
\sim 3 \times 10^{-5}~~.
\eeq
For the LMA solution ($\Delta m^2 \sim 
2 \times 10^{-5}$ eV$^2$) this ratio is even smaller :
$\ell_\nu / L \sim 10^{-5}$. For the low $\Delta m^2$ solution
($\Delta m^2 \sim 10^{-7}$ eV$^2$) this ratio is
$\ell_\nu / L \sim 1.5 \times 10^{-3}$, still quite small.

 From (\ref{phase}), we find that the phase $\varphi$ varies with the
energy as
\beq
\Delta \varphi = 
\frac{L \Delta m^2}{2 E} \frac{\Delta E}{E} =
2 \pi \frac{L}{\ell_\nu} \cdot
\frac{\Delta E}{E} ~~.
\eeq
Therefore, 
the change in energy as small as
$$\frac{\Delta E}{E} \approx \frac{\ell_\nu}{L}$$
leads to the change in the oscillation phase by $2\pi$.
As a consequence, the integration over even small intervals
of neutrino energy will
lead to the averaging of oscillations and the 
observed effect 
will be equivalent to that due to an incoherent mixture of
$\nu_1$ and $\nu_2$.

\bigskip

{\bf 3.}
Let us consider this in more details. Denoting the amplitudes of
$\nu_1 \to \nu_e$ and $\nu_2 \to \nu_e$ transitions inside
the earth by $b_{e1}$ and $b_{e2}$ respectively, we can write the amplitude
for finding $\nu_e$ in the detector as
\beq
\langle \nu_e | \nu_D \rangle 
= a_1~ b_{e1} + a_2~ b_{e2} ~e^{i(\phi_S + \varphi)}~~,
\eeq
then the probability of observing  $\nu_e$ at the detector is
\beq
P = a_1^2 ~|b_{e1}|^2 + a_2^2 ~|b_{e2}|^2 +
2 \mbox{ Re}[ a_1~ a_2~ b_{e1}^* ~ b_{e2} ~e^{i(\phi_S + \varphi)}]~~.
\eeq
Introducing $P_{2e} \equiv |b_{e2}|^2$ and using (\ref{a2def}),
we get
\beq
P = \frac{P_\odot - s^2 + P_{2e} (1-2 P_\odot)}{\cos 2\theta} 
+ P_{coh}(E) ~~,
\label{netprob}
\eeq
where
\beq
P_{coh} \equiv \frac{1}{\cos{2\theta}} \sqrt{(P_\odot - s^2) (c^2 - P_\odot) 
P_{2e} (1 - P_{2e})} \; \cos(\phi_S + \phi_E + \varphi)~~,
\eeq
with $\phi_E = \mbox{ Arg}(b_{e1}^* b_{e2})$. The first term in 
(\ref{netprob}) represents the incoherent
part of the probability and it coincides with the one used in 
the literature \cite{ms87,baltz,daynight}. The second term $P_{coh}$ 
corresponds to the interference contribution, 
which appears in the case of
coherence.

Let us consider the observable effects of $P_{coh}$. In the case 
of the SK detector, the number of events due to this ``coherent''
term is given by
\beq
N_e(E_e) \propto \int dE'_e  f(E_e, E'_e) \ \int_{E'_e}^{\infty}
dE_\nu \frac{d\sigma (E_\nu, E'_e)}{dE_e} \ 
F(E_\nu) P_{coh}(E_\nu)~~,
\eeq
where $F(E_\nu)$ is the original flux of neutrinos
at the detector without oscillations and 
$f(E_e, E'_e)$ is the (electron) energy resolution function 
of the detector.

Let us consider the integration over the neutrino energy:
\beq
I \equiv \int dE_\nu \cdot K(E_\nu) \cdot \cos[\phi_S(E_\nu) + \phi_E(E_\nu) + 
\varphi(E_\nu)]~~,
\label{nu-int}
\eeq
where
\beq
K(E_\nu) \equiv \frac{d\sigma (E_\nu, E'_e)}{dE_e} \ 
F(E_\nu) \frac{1}{\cos 2\theta} \sqrt{(P_\odot - s^2) (c^2 - P_\odot) 
P_{2e} (1 - P_{2e})} ~~.
\label{k}
\eeq
The crucial point is that 
$K(E_\nu)$ is a slowly varying function of energy as compared to 
$\varphi (E_\nu)$. Indeed,
the typical scale of the change of
cross-section, flux and probability in (\ref{k}) is
${\cal O}(\Delta E / E) \sim {\cal O}(1)$.
In particular, 
the variation in the probability 
$P_{2e}$ is of the order of 1 over an interval of
$\Delta E/E \sim \ell_\nu/(2 R_E)$, where
$R_E$ is the radius of the earth, and the latter ratio is
$\sim 1$.

The integration of the smooth function $K(E_\nu)$ 
with the rapidly oscillating function  
$\cos(\phi_S + \phi_E + \varphi)$ gives a result
very close to zero. Indeed, the 
integration in (\ref{nu-int})
can be performed first over small intervals $\Delta E$
and then the results can be summed: 
\beq
I \approx \sum_{k=0}^{k_{max}} 
\int_{E_0 + k \Delta E}^{E_0 + (k+1)\Delta E} dE_\nu K(E_\nu) \cdot
\cos[\phi_S(E_\nu) + \phi_E(E_\nu) + \varphi(E_\nu)]~~,
\eeq
where $E_0 = E'_e$. 
Choosing the interval $\Delta E$ such that 
$ \frac{\ell_\nu}{L} \ll \frac{\Delta E}{E} \ll 1$,
so that the change of $K(E)$ over this interval
can be neglected,
we can write
\barr
I & \approx & \sum_{k} K(E + k \Delta E) 
\int_{\Delta E} dE ~ \cos(\phi_S + \phi_E + \varphi)
\nonumber \\
& \approx  &  \int dE' ~ K(E') \frac{1}{\Delta E}
\int_{\Delta E} dE ~ \cos(\phi_S + \phi_E + \varphi)
\nonumber \\
& \approx & \frac{1}{2\pi} \frac{\overline{E}}{\Delta E}
\frac{\ell_\nu}{L} \cdot
\int dE' ~ K(E')~~,
\earr
where $\overline{E}$ is the typical energy in the interval 
of integration. 
The contribution of the coherent part of the integral
in comparison with the incoherent part can thus be estimated
as
\beq
\frac{N^{coh}}{N^{incoh}} \sim 
 \frac{1}{2\pi} \frac{\overline{E}}{\Delta E} \frac{\ell_\nu}{L}~.
\label{estimate}
\eeq
Taking $\frac{\Delta E}{\overline{E}} \sim 0.1$ we conclude, that 
to avoid the averaging, one should measure the energy of
the {\it neutrino} with an accuracy of better than $\sim 10^{-5}$,
which looks practically impossible. 

In the case of the LMA solution, the Boron
neutrino appears at the surface of the sun as a mass 
eigenstate $\nu_2$, which corresponds to $a_1 = 0$, and 
the regeneration in the earth is described by
$P_{2e}$ only, without any additional interference
effects.

In the case of the low $\Delta m^2$ solution, the 
high energy part of the Boron 
neutrino spectrum can be at the non-adiabatic edge, which means that
$a_1 \neq 0$, though it is rather small. 
Again, we need to measure the energy   
with an accuracy better than $\ell_\nu/ L \sim 10^{-3}$
to disentangle the coherence effects.

\bigskip

{\bf 4.}
Notice that even for a fixed energy, an ``averaging''
occurs due to our lack of knowledge about
the neutrino production point in the production region.
The size of the production region for Boron neutrinos 
inside the sun is nearly $ 0.1 R_\odot = 7 \cdot
10^9$ cm, an order of magnitude larger than the oscillation
length in vacuum.

One can estimate the averaging effects due to the production
of neutrinos from different points of the solar disc in the 
following way. 
For a small mixing angle, the transitions occur in a 
very thin resonance layer. Before the neutrinos enter 
this layer, their oscillations are strongly suppressed 
due to the large density, and after they exit the layer,
one can 
consider just vacuum oscillations. For the SMA
solution, the resonance layer is situated near
$R \sim 0.2 R_\odot$. The difference in the path length 
of neutrinos produced in different points of the solar disc
after they have crossed the resonance layer
is thus $\Delta L \sim 0.1 R_\odot$ and the averaging occurs since
\beq
\Delta \varphi = \frac{\Delta m^2}{2 E} \Delta L
= 2 \pi \frac{\Delta L}{\ell_\nu}
\eeq
and 
$\Delta L/ \ell_\nu \gg 1$ for the SMA and LMA solutions.
For the small $\Delta m^2$ solution, the effect due to
this averaging will be small. 
	
Note that in this approach, no averaging occurs 
due to the production of neutrinos along the line
of sight.

\medskip

Another origin of averaging is related to the eccentricity 
of the orbit of the earth.
Indeed, $\varphi$ is also a function of the time
$t$ of the year, with 
\beq
\Delta \varphi \sim \frac{\Delta L}{\ell_\nu} 
\left(\frac{\Delta t}{\mbox{1 year}}\right)~~,
\eeq
where $\Delta L$ is the variation in the distance 
from the sun to the
earth over one year. 
$\Delta L \approx 5 \times 10^{11}$ cm
so that $ \Delta L / \ell_\nu \sim 10^{3}$
for the SMA and LMA solutions. 
The integration over time $t$ of the year is then
reduced to the integration over the rapidly oscillating
function
$$
\int dt \cos[\phi_S + \phi_E + \varphi(t)]~~.
$$
The integration of the number of events over a period of
more than one day already leads to the averaging out of the
interference term (for the SMA and LMA solutions). 
If the events are binned in time
with $\Delta t \lsim (\ell_\nu / \Delta L)$ year, this averaging
can be avoided.

\bigskip

{\bf 5.}
The state which arrives at the earth
can be written in terms of flavor eigenstates as
\barr
\nu_E =  \frac{1}{\sqrt{\cos 2\theta}} & \times &
\left[ (c \sqrt{P_\odot - s^2} + s \sqrt{c^2 - P_\odot} \ e^{i(\phi_S + \varphi)}) ~\nu_e \right.
\nonumber \\
& & \left. + (s \sqrt{P_\odot - s^2} - c \sqrt{c^2 - P_\odot} \ e^{i(\phi_S + \varphi)}) ~\nu_\mu \right]~~.
\label{correct}
\earr
In the paper \cite{sehgal}, the state which arrives at the earth 
has been taken as
\beq
\nu_E = \sqrt{P_\odot} ~\nu_e + \sqrt{1-P_\odot} ~\nu_\mu ~~,
\eeq
so that the oscillation effects on the way from the sun to
the earth have been overlooked. Moreover, the phase between
the flavor eigenstates $\nu_e$ and $\nu_\mu$ has been fixed
independent of the production point and the energy
of the neutrino. This coincides with the correct equation
(\ref{correct}), only with either 
(i) $s=0$ and $\phi_S + \varphi = \pi$, or
(ii) $P_\odot = s^2$, which corresponds to the case when
the transitions inside the sun are completely adiabatic and
the state arriving at the earth is pure $\nu_2$.

For a correct description, the coherence effects should be 
discussed in terms of the mass eigenstates.

\bigskip

{\bf 6.}
To summarize, 
the relative phase between the neutrino mass eigenstates
arriving at the earth depends strongly on (i) the energy of
the neutrino, (ii) its production point and 
(iii) the earth-sun distance.
With a finite uncertainty in the measurements
of these quantities, the coherence effects due to this
relative phase get averaged out. 
In order to be able to observe these coherence effects,
one needs to  
(i) measure the energy of the neutrino
to an accuracy of $\Delta E / E \lsim 10^{-5}$,
(ii) determine the direction of the neutrino to an accuracy 
of better than 1/10$^{\mbox{th}} $ the size of the disc of the sun
and in addition, have 
some detailed knowledge about the production region, and  
(iii) bin the events in time with $\Delta t$ of the order of
a few hours.

	In all the practical cases, one can use the expression
(\ref{netprob}) without the $P_{coh}$ term, 
which corresponds to an incoherent mixture
of the mass eigenstates. The calculations of the 
regeneration effect which neglect the coherence
are then valid.

\end{document}